\documentclass[onecolumn,showpacs,preprintnumbers,amsmath,amssymb]{revtex4}
\usepackage[utf8]{inputenc}
\usepackage[english]{babel}

\usepackage{setspace}
\usepackage{booktabs}
\usepackage{multirow}
\usepackage{amsmath,amssymb}
\usepackage{qcircuit}
\usepackage{lipsum}
\usepackage{natbib}
\usepackage{braket}
\usepackage{stfloats}
\usepackage{graphicx}
\usepackage{subcaption} 
\usepackage{float}
\usepackage{physics}
\usepackage[figurename=Fig., justification=RaggedRight]{caption}
\usepackage{amsmath}
\usepackage{xcolor}
\usepackage{hyperref}

\begin{document} 

\title{Measurement induced phase transition in the central spin model: second Rényi entropy  in dual space approach}
\author{V. V. Belov$^{1,2}$, W. V. Pogosov$^{1,3,4,2}$ }
\affiliation{$^1$Dukhov Research Institute of Automatics (VNIIA), Moscow, 127030, Russia}
\affiliation{$^2$HSE University, Moscow, Russia}
\affiliation{$^3$Advanced Mesoscience and Nanotechnology Centre, Moscow Institute of Physics and Technology (MIPT), Dolgoprudny, 141700, Russia}
\affiliation{$^4$Institute for Theoretical and Applied Electrodynamics, Russian Academy of Sciences, Moscow, 125412, Russia}

\begin{abstract}

We conduct a numerical investigation of the dynamics of the central spin model in the presence of measurement processes. This model holds promise for experimental exploration due to its topology, which facilitates the natural distinction of a central particle and the quantum bath as different subsystems, allowing for the examination of entanglement phase transitions. To characterize the measurement-induced phase transition in this system, we employ a recently developed method based on second Rényi entropy in dual space. Our simulations account for decoherence, energy relaxation, and gate errors. We determine critical measurement rates and demonstrate that they significantly differ from those predicted by a simple approach based on mutual entropy.
\end{abstract}

\maketitle

\section{Introduction}

In recent years, there has been a surge of interest in exploring novel methodologies to mitigate quantum errors, with particular emphasis on the nascent domain of measurement-induced phase transition (MIPT).
 This emerging field offers promising avenues for error suppression in quantum systems. At its core, MIPT operates on a fundamental principle: while unitary operations generally enhance the entanglement degree of a system, measurements, conversely, tend to diminish it. For instance, the entanglement of two qubits can be achieved through unitary operations; however, the act of measuring one of the qubits results in the decoupling of their states.
 Let us stress that the term phase transition is used here with a different meaning compared to usual phase transitions (continuous or discontinuous ones), since  MIPT occurs in the space of quantum information and entanglement.

The primary inquiries in the realm of MIPT revolve around elucidating the behavior of systems comprising numerous qubits subjected to entangling unitary operations and random quantum measurements. The interplay between entangling unitary evolution and disentangling measurements can give rise to highly intricate dynamics, including phase transitions between different states of entanglement. These transitions depend upon factors such as the frequency or intensity of measurements taken \cite{skinner2023lecture,Skinner_2019,Li_2018,Gullans_2020,Granet_2023,Mazzucchi_2016,Li_2019,Chan_2019,Szyniszewski_2019,52866,Poboiko_2024}. At low measurement rates during the evolution from some initial state, an entanglement phase predominates, characterized by an initial decrease in entanglement due to measurements followed by stabilization at a nonzero level. Conversely, exceeding a critical measurement rate leads to a transition to the disentangling phase, where entanglement exponentially diminishes over time. Recent experimental investigations of MIPT have been conducted using systems comprising superconducting qubits \cite{Niknam_2021} and ion qubits \cite{Noel_2022}. The existence of MIPT was initially elucidated through approaches based on mutual entropy or Von Neumann entropy \cite{Skinner_2019,Gullans_2020}. However, recent studies \cite{Zhou_2021,Buchhold_2021,genLind} have introduced a new method based on Rényi entropy in dual space, which offers clear advantages by providing a more direct measure of entanglement degree.

In this paper, we investigate the dynamics of the central spin model \cite{Ashida_2019, Zhukov_2018} utilizing an approach based on second Rényi entropy \cite{Zhou_2021,Buchhold_2021,genLind} with postselection derived from the generalized Lindblad equation in dual space. The central spin model is well-suited for experimental exploration, as it comprises a single spin surrounded by a quantum bath of spins, with the central spin serving as a natural subsystem for determining entanglement entropy. Moreover, the symmetry of the central spin model greatly simplifies the theoretical analysis, allowing a larger number of qubits to be explored in numerical calculations compared to other spin models. We use both the simple mutual entropy approach and the second Rényi entropy approach as an indicator of phase transitions when analyzing the entanglement between the central spin (qubit) and the rest of the system. Our investigations also consider the effects of decoherence, energy relaxation, and gate errors, which are inevitable in real-world experiments. We assess their influence on the critical measurement rate and demonstrate that typical values of these parameters, characteristic of modern experimental setups, do not significantly suppress this quantity.

The paper is organized as follows. Section II defines the system under study. Section III formulates the method based on mutual entropy. Section IV presents an advanced approach based on second Rényi entropy. Section V analyzes the impact of gate errors, decoherence, and energy dissipation. We conclude in Section VI.

\section{Hamiltonian and preliminaries}

\begin{figure}[H]
    \centering    \includegraphics[width=0.4\textwidth]{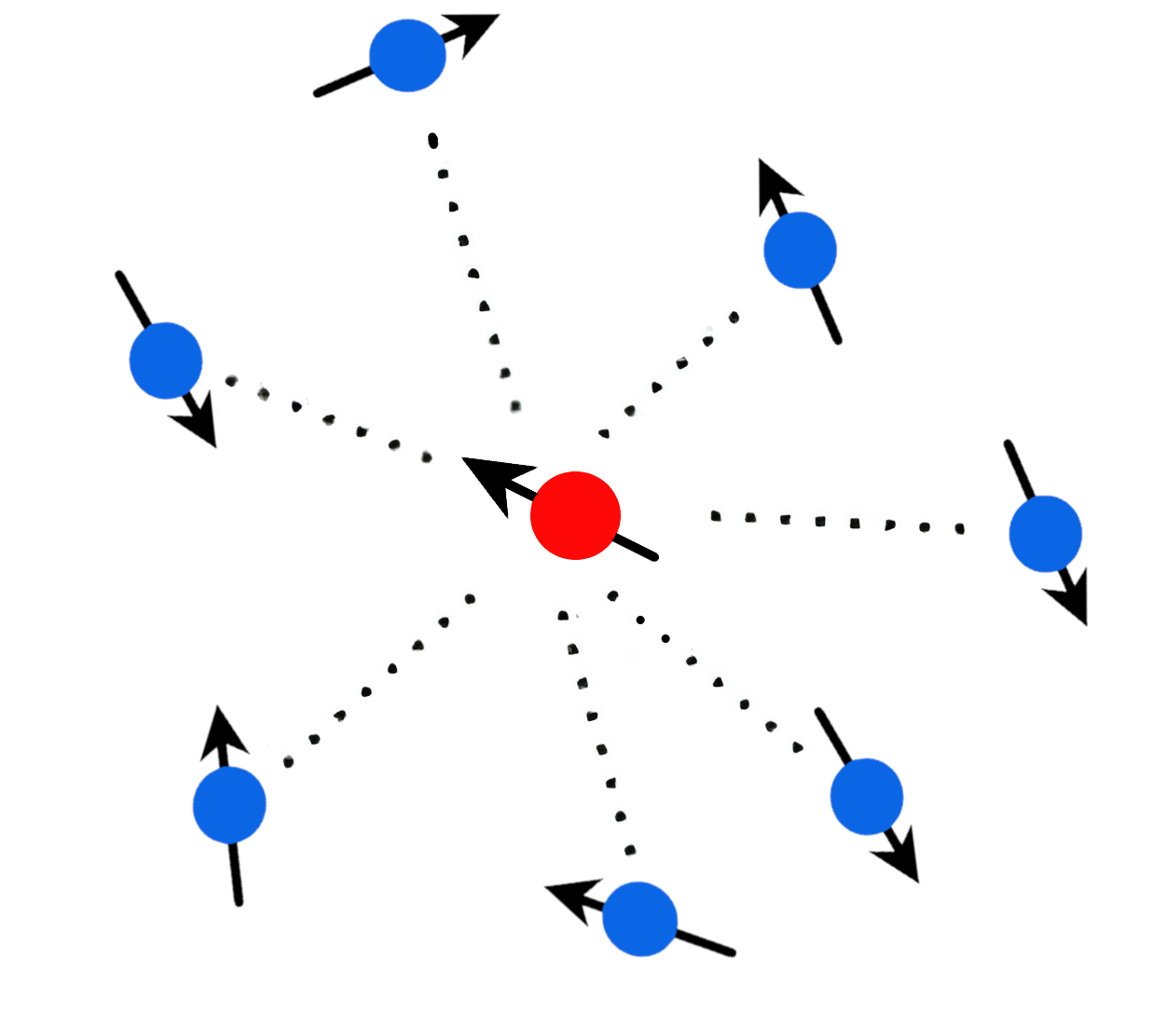}
    \caption{Schematic image of the system described by the central spin model. Central $1/2$-spin particle is interacting with the quantum bath of other spin-$1/2$ particles, which do not interact among themselves.}
    \label{fig:spinmodel}
\end{figure}

We consider a central spin model  \cite{Ashida_2019, Zhukov_2018}, which describes a spin system  shown schematically in Fig. \ref{fig:spinmodel}. The spin (qubit) system is represented by a central spin interacting with the quantum "bath" of other spins, which do not interact directly with each other. The system Hamiltonian is given by
    \begin{equation} \label{eq:6}
        {H} = {g} {\sum_{j = 1} ^{L} (\sigma_{c}^{+} \sigma_{j} ^{-} + \sigma_{c} ^{-} \sigma_{j} ^{+})} + \epsilon_c \sigma_{c,z}  + {\sum_{j=1}^{L} \epsilon_j \sigma_{j,z}},
    \end{equation}
where $\sigma_{j,z}$ and $\sigma^{\pm} _j$ are Pauli operators associated with particles of the bath, while $\sigma_{c,z}$ and $\sigma^{\pm} _c$ refer to the central spin; $\epsilon_j$ and $\epsilon_c$ are excitation energies of spins of the bath, whereas $g$ is the interaction constant between the central spin and each spin of the bath, $L$ is the number of spins in the bath. Hereafter we assume that $\epsilon_j=\epsilon_c$ for all $j$. 

We switch to the rotating frame, in which there is only one energy scale associated with $g$. Then, we introduce a dimensionless system of units, in which the energy is measured in terms of $g$, while time is measured in terms of $1/g$.
\qquad

\section{Phase transition model based on mutual entropy}
Let us start by examining the description of phase transition dynamics based on the concept of mutual entropy, which is technically simpler compared to the approach using Rényi entropy. In practical applications, measurements can be viewed as a quantum channel ${\xi (\rho_{SE}(0),t)}$ defined as \cite{CRESSER2006352,wiseman_milburn_2009,genLind}:
    \begin{equation} \label{eq:1}
        {\rho_S (t)} = { \Tr _{E} (U_{SE} (t) \rho_{SE} (0)  U_{SE} ^{\dag} (t)} = {\xi (\rho_{SE}(0),t)},
    \end{equation}
where $\rho_S$ refers to density matrix of the system under the study, $\rho_E$ is the density matrix of the medium in which measurements are performed, and $U_{SE}$ is the system plus environment evolution operator.

Next, we obtain the generalized Lindblad equation, which is used to describe the dynamics. Of crucial importance is the Kraus representation, which enables us to describe the system's state evolution through a set of operators known as Kraus operators. The Kraus representation establishes a relationship between the density matrix at time $t + \delta t$ and the density matrix at time $t$. It can be expressed as:
    \begin{equation} \label{eq:2}
        {\rho_s (t + \delta t)} = {\sum_k \hat{A_k} \rho_s (t) \hat{A_k ^{\dag}}} = {\xi (\rho_s)},
    \end{equation}
where $\hat{A_k}$ are Kraus operators. Thus, the dynamics arise from the action of the quantum channel, which ensures fundamental physical principles such as probability conservation.

The expression for the density matrix at $t + \delta t$ is given by \cite{CRESSER2006352,wiseman_milburn_2009,genLind}:
\begin{equation} \label{eq:3}
{\rho_s (t + \delta t)} = {\gamma \delta t \sum_k {\hat{A_k} \rho_s (t) \hat{A_k} ^{\dag}} + (1-\gamma \delta t) (\rho_s (t) - i[{H},\rho_s]\delta t) + O(\delta t^2)},
\end{equation}
where $\gamma$ represents the measurement degree. If each qubit is measured sequentially no more than once, $\gamma$ can be interpreted as the probability of qubit measurement per unit of time. The first term on the right-hand side of Eq. (\ref{eq:3}) describes the effect of measurement, while the second term is responsible for the unitary evolution. Specifically, since measurements occur with a probability of $\gamma \delta t$, it implies that with the opposite probability $(1-\gamma \delta t)$, the system will undergo the usual unitary evolution.
We use the Kraus operators in the standard form
\begin{equation}
\begin{aligned}
    A_0 &= 1 + \Delta t (L_0 - i K), \\
    A_k &= L_k \sqrt{\Delta t}, \quad t > 0,
\end{aligned}
\label{eq:Kraus}
\end{equation}
where $L_0$ and $K$ are Hermite operators. Substituting $A_0$ and $A_k$ in $\sum_k {\hat{A_k} \rho_s (t) \hat{A_k} ^{\dag}}$ of the right-hand side of Eq. \eqref{eq:3} we obtain:
\begin{equation}
\begin{aligned}
    \sum_{k=0} {\hat{A_k} \rho_s \hat{A_k} ^{\dag}} = {\hat{A_0} \rho_s \hat{A_0} ^{\dag}} + \sum_{k \neq 0} {\hat{A_k} \rho_s  \hat{A_k} ^{\dag}} = (1 +\delta t(L_0 - iK))\rho _s (1+\delta t(L_0 +iK)) + \delta t \sum_{k \neq 0} {\hat{L_k} \rho_s  \hat{L_k} ^{\dag}} =  \\
    \rho_s + \delta t (L_0 -iK) \rho_s + \rho_s \delta t( L_0 +iK) + \sum_{k \neq 0} {\hat{L_k} \rho_s  \hat{L_k} ^{\dag}} = \rho_s + \delta t(\{ L_0,\rho_s\} + i [ \rho_s, K] ) + \sum_{k \neq 0} {\hat{L_k} \rho_s  \hat{L_k} ^{\dag}}.
\end{aligned}
\label{eq:Kraus2}
\end{equation}

We make a replacement $K \rightarrow H$ and assume that $L_0 = - \frac{1}{2} \sum_{k = 1} \hat{L_k} ^{\dag} \hat{L_k}$ (from the condition of keeping a trace). Then from Eq. \eqref{eq:3} in the limit $\delta t \rightarrow 0$ we obtain the generalized Lindblad equation \cite{simple,Cao_2017}:
\begin{equation} \label{eq:4}
{\pdv{ \rho}{ t}} = {-[{H},\rho(t)] + \gamma \sum_{a = 1} ^{L+1} [\hat{L_a} \rho(t) \hat{L_a} ^{\dag} - \frac{1}{2} \lbrace \hat{L_a}^{\dag} ,\hat{L_a} \rho (t) \rbrace]},
\end{equation}
where $L_a$ represents the Lindblad operators or "jump" operators. If measurements are not performed on all qubits of the system, it is necessary to ensure trace preservation and Hermitianity for the Lindblad equation to be correct. If the measurements are projective and complete, we have:
\begin{equation}
\begin{aligned}
\Tr(\frac{d \rho}{dt}) = 0 =  \Tr(\{ \hat{L_0} , \rho \}) + \Tr(\sum_a ^{L+1}{\hat{L_a} \rho \hat{L_a}^{\dag}}).
\end{aligned}
\label{eq:chek1}
\end{equation}
It can be seen that in order for Eq. \eqref{eq:chek1} to be correct for incomplete measurements, it is necessary to add a normalization factor to the second term of the right-hand side of this equation.
Incomplete measurements are required because otherwise, postselection would be necessary after all experiments, leading to an exponentially increasing number of measurements \cite{friedman2023measurementinduced}. To ensure that Eq. \eqref{eq:4} preserves the trace of the matrix, we normalize the second term on the right-hand side of this equation  and obtain \cite{CRESSER2006352,wiseman_milburn_2009,genLind}:
\begin{equation} \label{eq:5}
{\pdv{ \rho}{ t}} = {-[{H},\rho(t)] + \gamma \sum_{a = 1} ^m \frac{\hat{L_a} \rho(t) \hat{L_a} ^{\dag}}{\text{Tr}(\sum_{b=1} ^m \hat{L_b} \rho(t) \hat{L_b} ^{\dag})} - \frac{\gamma}{2} \sum_{a=1} ^n \lbrace \hat{L_a}^{\dag} ,\hat{L_a} \rho (t) \rbrace},
\end{equation}
where $m$ ($m < L+1$) represents the number of qubits that are measured.

Understanding the evolution of the density matrix enables us to calculate the entropy, which serves as an indicator of the system's phase. Investigating entanglement between two subsystems, namely the single qubit and the rest of the qubit system, is particularly advantageous for experimentally reproducing and analyzing the presence or absence of a MIPT \cite{Noel_2022}. A simple approach in determining the entanglement of a system is based on a concept of a mutual entropy. Namely, one calculates the mutual entropy of two subsystems (central qubit + surrounding qubits) as $S_{mutual} = S(\rho_a) +S(\rho_b) - S(\rho_{ab})$. The components here are calculated, for example, using the von Neumann entropy of the corresponding subsystem or of the whole system as $S(\rho) = -Tr(\rho \log \rho)$. This method was, for instance, used in  Refs. \cite{PhysRevResearch.2.013022,enRen}. It is an approach that gives only an upper bound estimate of the residual entropy in the system. We discuss this issue in a more detail in Section IV.
 \qquad

\begin{figure}[H]
\centering
\includegraphics[width=0.6\textwidth]{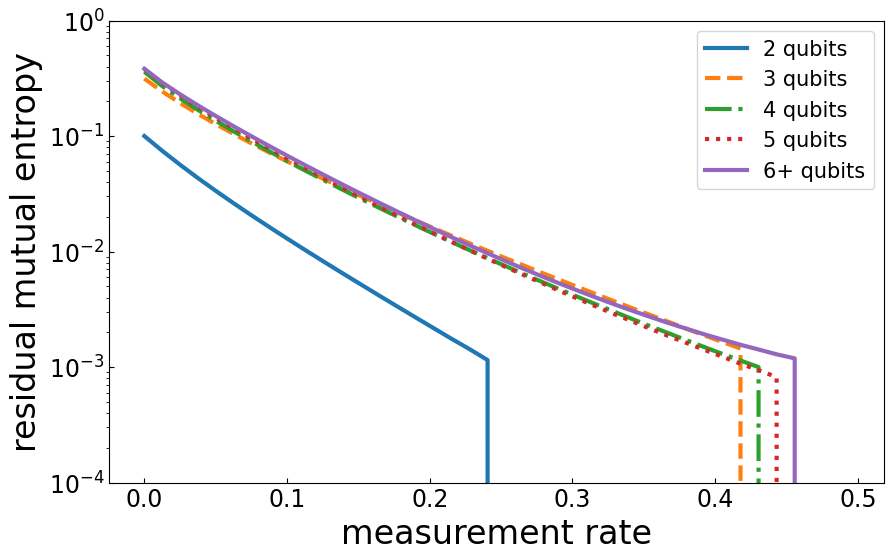}
\caption{Residual mutual entropy after the free evolution over $t=15$ (in dimensionless units) as a function of the measurement rate in systems with varying numbers of qubits.}
\label{fig:mutual-entropy}
\end{figure}

In the straightforward scenario where measurements are considered as instantaneous projections onto specific quantum states, Lindblad operators can be expressed as:
\begin{equation} \label{eq:7}
\begin{split}
{\hat{L}_{i,0}} = {\frac{1}{\sqrt{N}}| 0 \rangle_i \langle 0 |},
\\
{\hat{L}_{i,1}} = {\frac{1}{\sqrt{N}}| 1 \rangle_i \langle 1 |},
\end{split}
\end{equation}
where $i = 1,2,\ldots L+1$ and $N=2^{L+1}$. For more complex measurement scenarios, it may be necessary to employ advanced methods described in Refs. \cite{opensys,Scala_2008}.
In our study, the quantum system and its dynamics are modeled using the Qutip library \cite{JOHANSSON20121760,JOHANSSON20131234}.
Analyzing the entropy of a system involves studying its temporal evolution to determine whether the entropy has reached saturation (residual entropy) or has completely disappeared after a large amount of time. This approach allows us to assess the stability of the system state and the presence of phase transitions: if any entropy remains, we are in the entanglement phase, if not, we are in the disentangled phase. The value of the measurement rate at which the entanglement phase is observed as measurement rate is decreased is referred to as critical rate. For this analysis, the system's state after the free evolution for $t \gg 1$ (in dimensionless units) is chosen, as saturation is observed at this stage. In particular, in our calculations, we have chosen $t=15$ to probe the entanglement. Random mixed initial states are employed to observe the purification phase transition \cite{skinner2023lecture,Gullans_2020}. The results depicting mutual entropy versus measurement rate $\gamma$ for different numbers of qubits are presented in Fig. \ref{fig:mutual-entropy}. Beginning with a system comprising six qubits, the curves start to overlap, indicating that the system size no longer affects the critical measurement rate.

\begin{figure}[H]
\centering
\includegraphics[width=0.45\textwidth]{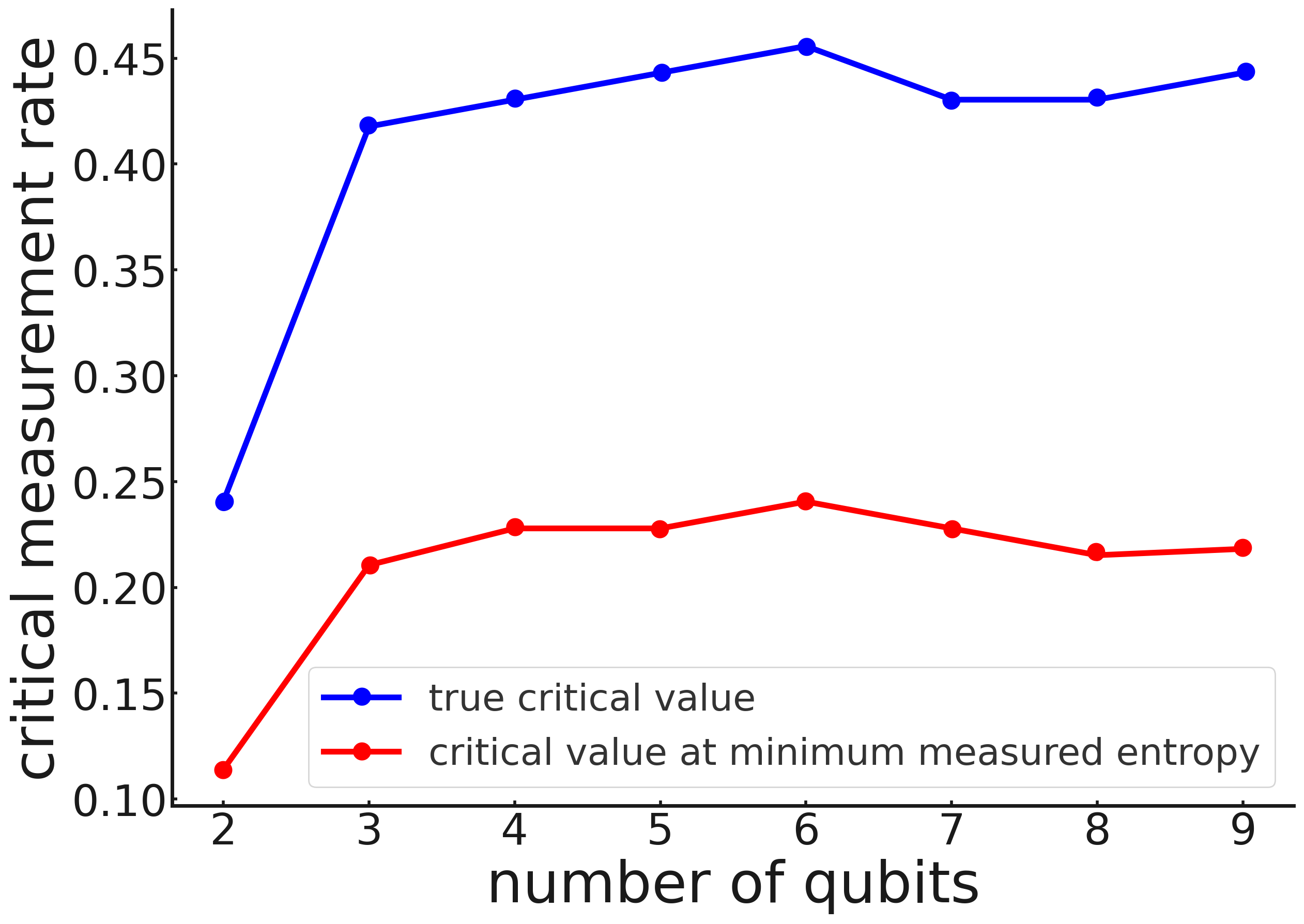}
\caption{Dependence of the critical value on the number of qubits using the approach based on mutual entropy. The upper graph illustrates the actual dependence of the critical value on the system size, while the lower one is calculated assuming that the minimum measurable entropy is approximately $10^{-3}$(that is, entropy values below $10^{-3}$ are considered to be 0). Solid curves serve as guides for eyes connecting discrete points.}
\label{fig:simple_speed}
\end{figure}

As depicted, the critical measurement rate corresponding to the transition to the entanglement state increases with the system size. This arises from the fact that in smaller systems, even a single qubit measurement can significantly diminish the level of entanglement. However, as the system size increases, it is anticipated that the impact of individual measurements on the overall entanglement will diminish, and the critical value should converge to a saturation limit, denoted as $P_{crit}$, which no longer depends on the system size and is approximately 0.22. Additionally, it is observed that starting from a system of nearly 9 qubits, the critical value reaches saturation. However, the deviation is sufficiently small to not warrant further investigation.

\section{Application of the generalized Lindblad equation in a dual system}

In the realm of quantum systems dynamics, it has become obvious that the naive approach used to analyze the phase information transition of entanglement does not always reflect all the subtleties within the system. This issue is discussed in a more detail in recent studies \cite{Zhou_2021,Buchhold_2021,genLind}, which also propose a new approach based on the Rényi entropy in a dual subsystem. We now briefly show why the method described in Section III gives only an upper bound for the critical measurement rate. 
The density matrix of the subsystem of the whole system can be expressed as
\begin{equation} \label{eq:matrix1}
\rho _ a = \sum _i {p_i  \Tr_b \rho_{ab_i}}.
\end{equation}
It can be seen that the total entropy consists of two contributions, which originate from the distribution of the measurement result itself $Tr_b \rho_{ab_i} =\rho _{a,i}$, responsible for entropy of entanglement, and the probability distribution to obtain this state $p_{i}$. For MIPT, we are interested only in the first contribution, while the second contribution must be discarded.
This can be done by averaging the probabilities of the various results and determining the new entropy of entanglement through the second  Rényi entropy, but without taking into account $p_i$. Thus, we obtain
\begin{equation} \label{eq:true_answer}
S^{'} _a = -\log(\sum_i \tilde{p}_i \Tr_a{\rho^2 _{a,i}}).
\end{equation}
where $\tilde{p_i} = \frac{p^2 _i}{\sum_{i'}p^2 _{i'}}$. In order to find entropy according to Eq. \eqref{eq:true_answer} we need to transform the original system \cite{Zhou_2021,Buchhold_2021,genLind}. For example, it is proposed in Refs. \cite{Zhou_2021,Buchhold_2021,genLind} to use the concept of a doubled system and the properties of Einstein-Podolsky-Rosen (EPR) pairs. Having done this, it is possible to correctly introduce the Rényi entropy, thereby bypassing the influence of the probability distribution of measurements on the second Rényi entropy. In this case, the original equation \eqref{eq:5} is transformed into the generalized Lindblad equation for a binary system as \cite{Zhou_2021,Buchhold_2021,genLind}:
    \begin{equation} \label{eq:8}
        {\pdv{ \rho_D}{ t}} = {-[{H_D},\rho_D(t)] + \gamma \sum_{a = 1} ^m \frac{\hat{L}_{a,L}  \hat{L}_{a,R}\rho_D(t) \hat{L}_{a,L} ^{\dag} \hat{L}_{a,R} ^{\dag}}{Tr(\sum_{b=1} ^m \hat{L}_{b,L}  \hat{L}_{b,R}\rho_D(t) \hat{L}_{b,L} ^{\dag} \hat{L}_{b,R} ^{\dag})} - \frac{\gamma}{2} \sum_{a,b=1} ^n \lbrace  \hat{L}_{a,L}  \hat{L}_{a,R} \hat{L}_{b,L}  \hat{L}_{b,R} , \rho_D (t) \rbrace},
    \end{equation}
    where $\rho_D = \rho \otimes \rho$, $H_D = H \otimes I + I \otimes H$, $A$ and $B$ are subsystems (in our case $A$ contains $L$ qubits, while $B$ is represented by a single qubit). Then the total calculated entropy, which can still be measured in experiments, is given by: 
    \begin{equation} \label{eq:9}
        {S_{D}} = {-\log_2(\Tr_{L_A,R_A} [X_A \Tr_{L_B,R_B} (\rho_D)])}.
    \end{equation}
where $X_A$ is a SWAP operator in subsystem $A$, it acts as $X \ket{a _L} \ket{b _R} = \ket{b _L} \ket{a _R}$.

\begin{figure}[H]
\centering
\includegraphics[width=0.6\textwidth]{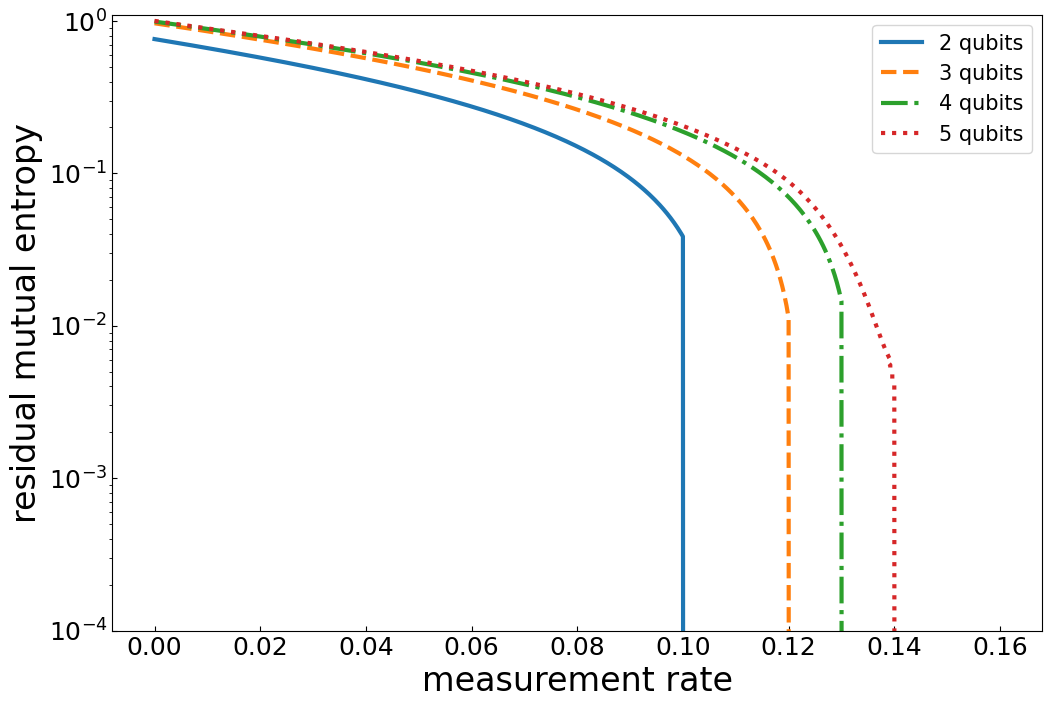}
\caption{Residual second Rényi entropy after the free evolution during the time $t=15$ (in dimensionless units) as a function of the measurement rate in systems with different numbers of qubits.}\label{fig:hard_sum}
\end{figure}


To model the dynamics of the doubled system, we once again employ the Qutip library \cite{JOHANSSON20121760,JOHANSSON20131234}. Simulations can be conducted for either a pure or mixed initial state of the system, with the choice of initial state determining different types of measurement-induced phase transitions (MIPT) \cite{skinner2023lecture,Gullans_2020}. As previously, we select a random mixed state as the initial state of the system to study the dynamical purification phase transition. The resulting dependencies of the residual entropy on the measurement rate are presented in Fig. \ref{fig:hard_sum}.

\begin{figure}[H]
\centering
\includegraphics[width=0.45\textwidth]{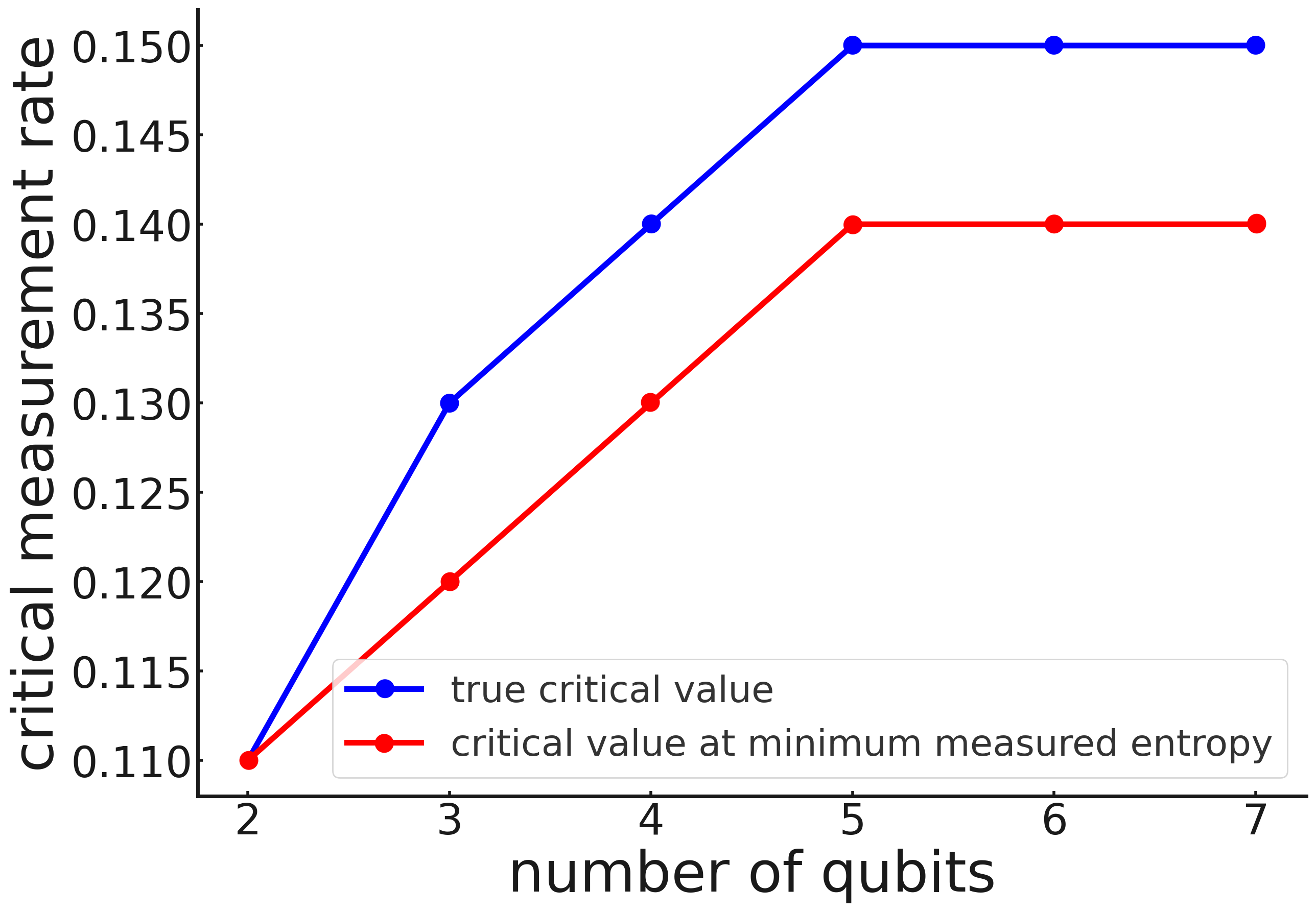}
\caption{Dependence of the critical measurement rate on the qubit number in a double system. The upper graph corresponds to the actual dependence of the critical value, while the lower one is calculated under the assumption that the minimum measurable entropy is approximately $10^{-3}$(that is, entropy values below $10^{-3}$ are considered to be 0). Solid curves serve as guides for eyes connecting discrete points.}
\label{fig:hard_speed}
\end{figure}

As evident from Figs. \ref{fig:simple_speed} and \ref{fig:hard_speed}, the second Rényi entropy yields critical values that differ from those obtained using the mutual entropy method. It stabilizes at a constant value with fewer qubits (as early as 5 qubits), unlike mutual entropy, which stabilizes at nearly 9 qubits. The critical value for mutual entropy, under the condition of bounded entropy measurement, is approximately $0.22$, whereas for the second Rényi entropy, it is approximately $0.14$. Thus, Rényi entropy and mutual entropy offer distinct approaches to estimating entanglement and information distribution within the system, resulting in varying critical value estimates. So we showed that a really simpler approach to modeling also takes into account the contribution that does not affect MIPT, at the same time using the method based on the second Rényi entropy is quite difficult since it requires an analysis of a binary system.
Thus, our conclusions in both cases align with the finding that entanglement phase transitions are possible in the central spin model. We were also able to examine how the description of MIPT differs when considering different entropies.

\section{The impact of gate errors, energy relaxation, and decoherence}

\begin{figure}[H]
\centering
\includegraphics[width=0.45\textwidth]{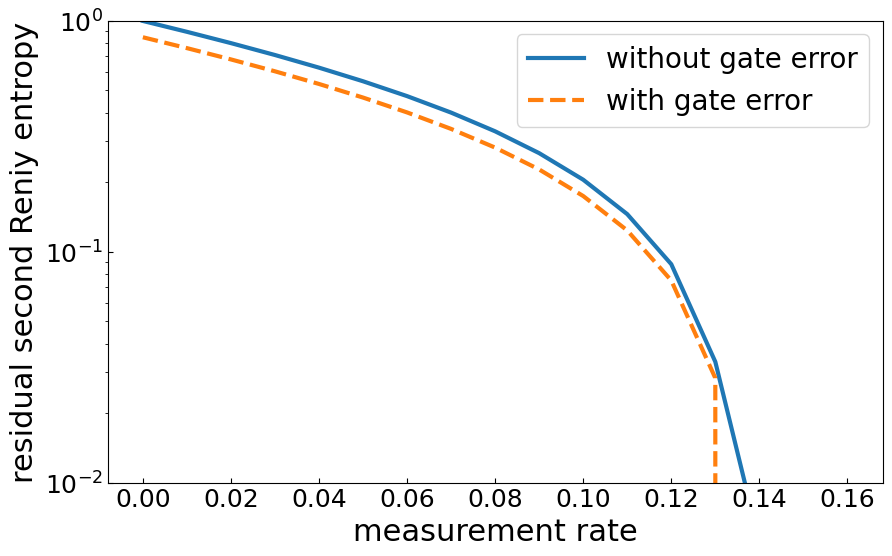}
\vspace{1.5 cm}
\includegraphics[width=0.45\textwidth]{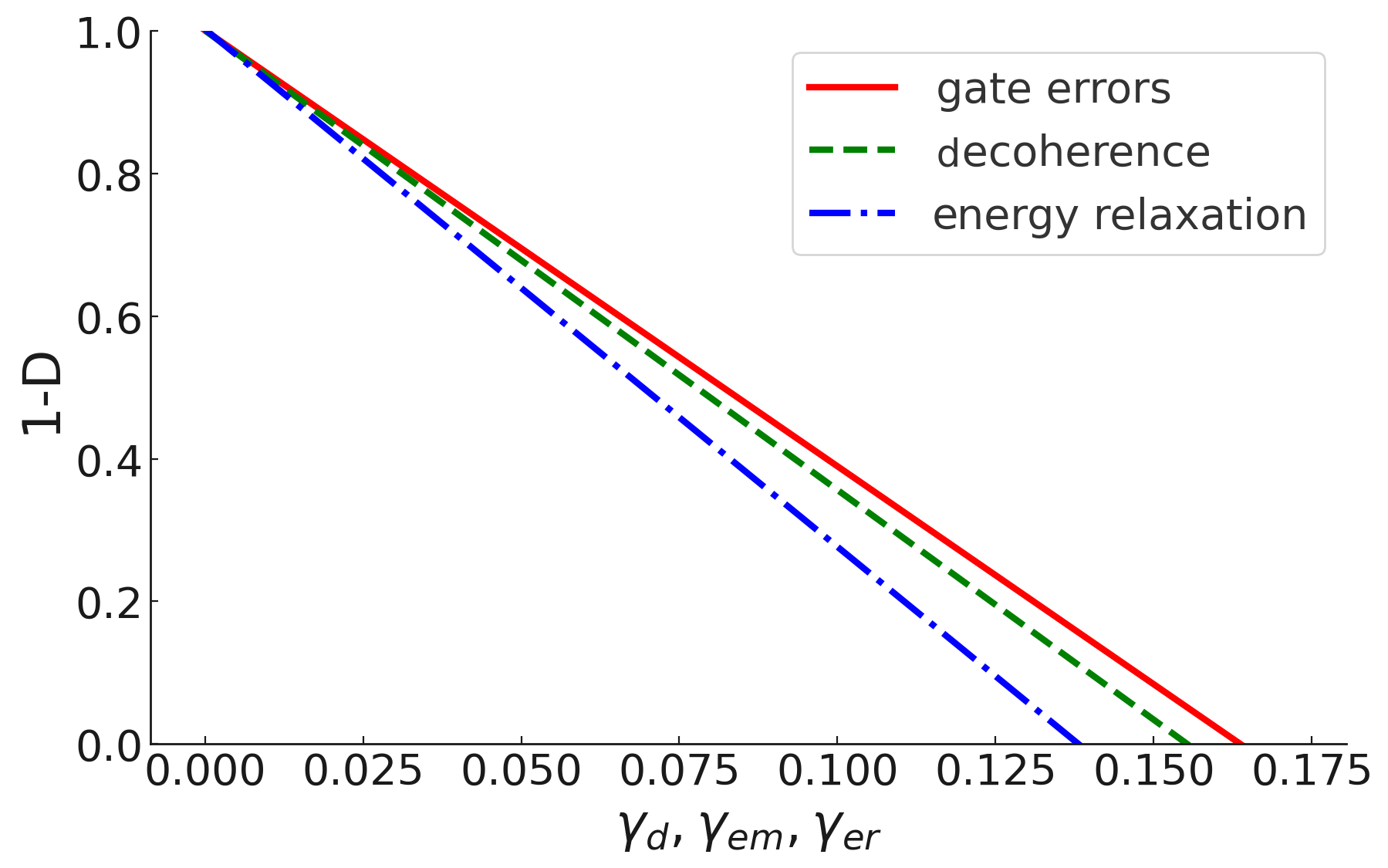}
\caption{(a) Residual second Rényi entropy as a function of measurement rate without and with gate errors. (b) $1-D$ as a function of energy relaxation rate, decoherence rate, and gate error rate.}
\label{fig:errors}
\end{figure}

Gate errors, energy relaxation, and decoherence can significantly influence the dynamics of quantum systems, impacting both the evolution of unitary entanglement and the accuracy of measurements. Understanding the criticality of these factors is essential, as MIPT is expected to be utilized for quantum error correction. Errors can lead to unexpected changes in the system's state, reducing the efficiency of entanglement evolution and decreasing measurement accuracy. Energy relaxation, on the other hand, can diminish the overall level of entanglement in the system, thereby affecting the observability of phase transitions.

In the following, we will focus solely on the second Rényi entropy, although the influence of these negative effects will be analogous for mutual entropy. In the simplest case of the Markov approximation, where the 'reservoir' relaxation is fast and the interaction is weak, all these effects can be described through additional Lindblad operators \cite{Tarnowski_2023,Howard_2006}. While more general methods are available \cite{Samach_2022,opensys}, the results remain practically unchanged. The Lindblad operators for each of the main negative processes are expressed in a standard form -- decoherence is described by $L_{rel} = \sqrt{\gamma_{d}} \sigma ^z$, energy relaxation by either $L_{em} = \sqrt{\gamma_{em}} \sigma ^{+}$ or $L_{abs} = \sqrt{\gamma_{abs}} \sigma ^{-}$, and gate errors by $L_{er} = \sqrt{\gamma_{er}} \sigma ^{y}$ or $\sqrt{\gamma_{er}} \sigma ^{x}$. We then substitute these additional Lindblad operators into Eq. \eqref{eq:8}. We simulate each of these effects separately, generating a new random mixed state many times and presenting the results in Fig. \ref{fig:errors}. Essentially, the effect of gate errors, energy relaxation, and decoherence is to shift all graphs in Fig. \ref{fig:hard_sum} to the left. Specifically, the critical value decreases due to these effects. Figure \ref{fig:errors} (a) illustrates an example of curve displacement when errors are added to the system. $D$ is the relative displacement defined as $ \frac{S_{D_1} - S_{D_2}}{S_{D_1}}$, where $S_{D_1}$ is residual entropy without errors, while $S_{D_2}$ is the same quantity in presence of errors. Figure \ref{fig:errors} (b) shows $1-D$ as a function of the energy relaxation rate, decoherence rate, and gate error rate. We observe that all three dependencies are linear, and the critical value vanishes at approximately the same value of these rates.

The simulations demonstrate that in experiments, we can achieve a regime where entanglement prevails, provided that $g$ is one or two orders of magnitude larger than $\gamma_i$. This requirement is realistic, for instance, for superconducting physical platform. We also note that the largest negative influence is exerted by the energy relaxation process, while the smallest is by quantum gate errors.

\section{Conclusions}

In this study, we examined MIPT for the central spin model, which characterizes a single two-level particle interacting with a quantum bath composed of other two-level systems. This model is pertinent for describing phenomena observed in quantum dots and defect centers in diamond. In the realm of MIPT, the central spin model holds promise for experimental quantum simulation using artificial quantum systems like superconducting qubits, owing to its topology that naturally segregates the central particle from its quantum bath.

We determined the critical measurement rate marking the transition from one phase to another using two approaches: (i) the conventional method based on mutual entropy and (ii) a recently proposed theoretical framework \cite{Zhou_2021,Buchhold_2021,genLind} based on the second Rényi entropy in dual space. Our findings revealed that both approaches exhibit similar tendencies, yet the second Rényi entropy approach yields a significantly smaller critical measurement rate.

Furthermore, we explored the impact of energy relaxation, decoherence, and quantum gate errors, all of which are inevitable in real-world experiments. Through numerical modeling, we assessed their effects and demonstrated the feasibility of observing MIPT in the central spin model with modern artificial quantum systems.

\section{Acknowledgments}\label{s:acknowledgments}

W. V. P. acknowledges support from the Grant of the Ministry of Science and Higher Education of the Russian Federation No. 075-15-2024-632 dated June 14, 2024.

\bibliography{auth}
\end{document}